\documentclass{agujournal2019}

\journalname{Geophysical Monograph 259}

\begin{document}

%
%


\title{Cold Ionospheric Ions in the Magnetosphere}

%
%




\authors{M. Andr\'e\affil{1}, S. Toledo-Redondo\affil{2}, and A. W. Yau\affil{3}}

\affiliation{1}{Swedish Institute of Space Physics, Uppsala, Sweden}
\affiliation{2}{Institut de Recherche en Astrophysique et Plan\'etologie, Universit\'e de Toulouse, CNRS, UPS, CNES, Toulouse, France.}
\affiliation{3}{University of Calgary, Calgary, Alberta, Canada}






\correspondingauthor{M. Andr\'e}{mats.andre@irfu.se}




\begin{keypoints}
\item Cold (eV) ionospheric ions are a major source of magnetospheric plasma
\item Ionospheric ions are important on large (MHD) scales
\item Cold ions are important on small (kinetic) scales
%
\end{keypoints}

\vspace{50mm}
{\Large An edited version of this paper was published by American Geophysical Union (2021).\\}
\vspace{5mm}
{André, M., S. Toledo-Redondo and A. W. Yau (2021) Cold Ionospheric Ions in the Magnetosphere, in Space Physics and Aeronomy Collection Volume 2: Magnetospheres in the Solar System, Geophysical Monograph 259, Chapter 15, doi: 10.1002/9781119815624.ch15}

%
%


\begin{abstract}
Cold (eV) ions of ionospheric origin are a major source of magnetospheric plasma. These ions dominate the number density for most of the volume of the magnetosphere, most of the time. 
This affects large-scale physics, including plasma temperature and pressure (and thus instability thresholds) and the Alfv\'en velocity (and thus energy transport with waves and the magnetic reconnection rate).
This also affects small-scale kinetic plasma physics, including wave generation and the Hall physics of reconnection. Ions escaping from the ionosphere also represent a significant fraction of the mass outflow from planet Earth.   
\end{abstract}

%
%

%


%
%
%
%

\section{Introduction: Cold Ions As Part of the Magnetosphere}

Cold ions of ionospheric origin are a major and important part of the terrestrial magnetosphere. Ions escaping the F-layer ionosphere of the Earth have an initial energy of less than 0.5 eV \cite{Kelley2009}. These ions may be energized to at least several keV but a major fraction never reaches energies beyond a few tens eV even after a long time in the magnetosphere. Here we focus on the effects and destiny of these initially cold ions. 

It was early suggested that ionospheric ions are a major source of magnetospheric plasma although this was hard to verify \cite{Chappell1980, Moore1984, Olsen1985, Chappell2015}. Cold positively charged ions are difficult to detect in low density plasmas since a sunlit metal object such a spacecraft becomes positively charged due to emissions of photoelectrons, hence repelling the ions. Using multiple techniques it has now become clear that cold ions are a common part of the magnetospheric plasma. These ions change the Alfv\'en velocity because they mass load the plasma (and thus change the energy transport with waves and the magnetic reconnection rate), and also the plasma temperature and pressure (and thus the threshold for Kelvin-Helmholtz instabilities). The cold ions also affect kinetic plasma physics, including wave generation and the Hall physics of magnetic reconnection. The initially cold ions also represent a significant fraction of the mass outflow from planet Earth.  

 In the following ions are considered cold when the thermal energy is less than tens of eV. The drift (bulk) energy can be higher; hundreds of eV is not unusual, often due to {\textbf{E} $\times$ \textbf{B}} drift. Ionospheric sources of cold ions include the high latitude cusp and cleft, the auroral region and the polar cap, and the mid- and low-latitude plasmasphere.

%

\section{Observations: Problems and Solutions}

A sunlit spacecraft in a low density plasma in Earth orbit becomes positively charged (tens of volts). Hence, positively charged ions at low energies will not reach the spacecraft and cannot directly be detected. Various techniques can be used to overcome this challenge.  

\subsection{Remote Sensing}

Cold ions can be detected as part of the total plasma content by using various remote sensing techniques. Actively transmitting groundbased ionosondes, and top-side sounding from a radio transmitter onboard a spacecraft, can be used to determine the total ionospheric density at a specific altitude \cite{Benson2010}. With groundbased radars and incoherent scatter radars, several parameters including plasma drift and ion and electron temperatures can be deduced \cite{Ogawa2009}. Instruments onboard spacecraft can detect EUV solar photons resonantly scattered from (usually minor species) He{\textsuperscript{+}} ions. Energetic neutral atoms (ENAs) produced by charge-exchange between magnetospheric ions and hydrogen atoms in the exosphere can travel in line-of-sight paths to a spacecraft and can be detected at energies at least down to tens of eV \cite{Sandel2003}. 

\subsection{Observations In Situ}
Detecting plasma in situ onboard a spacecraft using particle detectors such as electrostatic analyzers gives the opportunity to obtain details of the plasma properties but adds the problems caused by interaction of the spacecraft itself with the plasma. In the ionosphere and lower magnetosphere the density can be so high that a spacecraft potential can be close to zero or negative due to many impacting electrons. At altitudes of a few hundred km ion detectors can be used to study positive cold ions with energies of one or a few eV \cite{Shen2018}. Also, Langmuir probes can be used to determine electron and ion density, temperature and other parameters \cite{Brace2013}.

In a low density plasma at higher altitude, with a positively charged spacecraft, cold positive ions can still be observed. The spacecraft potential can be reduced by actively emitting a plasma cloud or positive ions, but often a potential of at least a few volts remains \cite{Moore1997, Torkar2016}. Also, during shorter periods when a spacecraft is temporarily in eclipse and hence negatively charged, cold positive ions can reach onboard detectors \cite{Seki2003}. An indirect method is to estimate the total plasma density from wave observations at the plasma frequency and subtract the hot ion density directly observed by particle detectors \cite{Sauvaud2001}. Another indirect method is to estimate the density from the spacecraft potential. This potential depends on the density and the electron temperature but can in low density magnetospheric plasma often be calibrated and used to estimate the total density \cite{Lybekk2012}. As an additional alternative method we note that cold positive ions in a supersonic flow can create a large enhanced wake behind a positively charged spacecraft. The wake will be filled with electrons, whose thermal energy is higher than the ram kinetic energy, in contrast to that of the ions. This will create a local wake electric field which can be observed and can be used to identify cold ion distributions. Using multiple instruments to observe also the geophysical electric field, the magnetic field and the spacecraft potential to estimate the density, the cold ion flux can be deduced \cite{Engwall2009a, Engwall2009b, Andre2015a}. To compare methods, we note that a mass spectrometer onboard a spacecraft can identify ion species but is affected by the spacecraft potential, while using the spacecraft potential and wake includes also the lowest energy ions but does not directly identify ion masses. Finally, many studies concentrate on initially cold ionospheric ions that have been heated (larger thermal energy) or that are drifting (e.g. {\textbf{E} $\times$ \textbf{B}} drifting) with low thermal but higher drift velocity. Here particle detectors on a slightly positively charged spacecraft can still be used. 

\section{Statistics and Effects at Large Scales}

\begin{figure*}
	\includegraphics[width=80 mm]{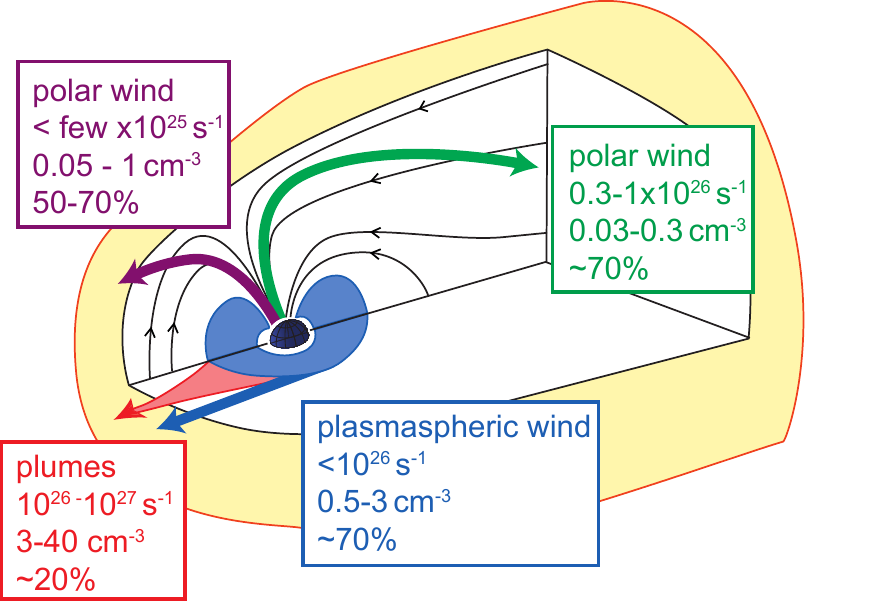}
    \caption{Overview of cold ion outflow. The sketch gives some major outflow paths, typical outflow rates and densities, and the approximate fraction of the time cold ions dominate the number density. Cold ions often dominate the density of the magnetosphere (from \citeA{AndreCully2012}). 
    }
    \label{fig:figure-Overview_GRL_2012}
\end{figure*}

Many studies show that ions of ionospheric origin are an important part of the magnetosphere \cite{Welling2015}, including heavier ions such as O$^+$ and He$^+$ \cite{Kronberg2014}. Several studies of cold upflowing ions are consistent with the overall sketch presented by \citeA{AndreCully2012} (Figure~\ref{fig:figure-Overview_GRL_2012}) although exact numbers depend on conditions during the study and on the measurement technique. In Figure~\ref{fig:figure-Overview_GRL_2012} the spacecraft wake technique was used for the polar regions (magnetotail lobes) and multiple detection techniques for the dayside magnetopause. We find that large regions of the magnetosphere are often dominated (in number density) by cold ionospheric plasma. 

For the low-density (nightside) lobes the ions often have energy that is so low that only the spacecraft wake method can be used for large statistical studies, \cite{Engwall2009a, Engwall2009b}. Using Cluster data covering nearly a full solar cycle it is clear that a major part of the time the cold ions dominate the density and upflow rate \cite{Andre2015a}. This rate depends on solar zenith angle and varies with season: solar illumination plays an important role in the state of the local ionosphere, which is a main factor for determining the outflow \cite{Maes2017}. Ions can stay rather cold as they flow outward in the lobes, or be energized to several eV by waves and centrifugal acceleration. This depends on the geophysical conditions, and also on outflow path and ion mass \cite{Yau2007, Nilsson2012}. Both more energetic ions from the auroral zone and cold ions from the polar cap can move into the lobes, with the upflow of O$^+$ increasing more with increasing EUV and geomagnetic activity than the upflow of H$^+$ \cite{MaggioloKistler2014, Slapak2017}.

The upper part of Table 1 shows estimates of upflow rates from high latitudes, including the cusp, cleft, auroral region and the polar cap, as determined by various statistical studies. This upflow depends on the solar EUV level (which varies with the solar cycle and affects the ionospheric density) and the magnetospheric disturbance level, as discussed in the individual references. The ion trajectories depend strongly on the convection pattern (the large-scale convection electric field) and the ions may not directly leave the magnetospheric system. 

When the lowest-energy ions (below a few eV) are included in the observations, using the Cluster wake method, or at rather low altitudes (where spacecraft charging is less of a problem), the measured upflow rate is of the order 10$^{26}$ ions/s. Using a combination of particle detectors and spacecraft potential control down to about +2 V at an altitude of 8 R$_E$ gives the same order of magnitude, but a significant fraction of the upflowing distribution is still often missed \cite{Su1998, Engwall2009a}. When the lowest energies are not included, the observed upflow rate can be one order of magnitude lower, indicating that most ions are not observed, see e.g. \citeA{Peterson2008}.

For the magnetospheric side of the (dayside) magnetopause there are statistical studies of ionospheric plasma using particle detectors on, e.g., Polar \cite{Chen2006}, THEMIS \cite{LeeAngelopoulos2014}, MMS \cite{Fuselier2017}, and multiple geosynchronous satellites with Magnetospheric Plasma Analyzers (MPA) \cite{BorovskyDenton2008}. Multiple observational methods using Cluster include the spacecraft wake technique, cold ions with substantial {\textbf{E} $\times$ \textbf{B}} drift, and comparison of total density from the plasma frequency with ion density from particle detectors \cite{AndreCully2012}. Remote sensing techniques on IMAGE used the solar EUV photons resonantly scattered by He$^+$ to estimate the total outflow \cite{SpasojevicSandel2010} and focused on high-density plasma plumes. These studies are consistent with direct (particle detector) observations of a persistent plasmaspheric wind from the plasmasphere to the magnetopause when the magnetospheric convection is not strong enough to form plasmaspheric drainage plumes \cite{Dandouras2013}. 

The lower part of Table 1 shows estimates of the outflow rate through the magnetopause.  For plasmaspheric plumes, the outflow rate can be at least a few times $10^{26}$ ions/s, but these plumes occur only about 10-20\% of the time \cite{AndreCully2012,Lee2016}. At other times a plasmaspheric wind is common, but with a lower outflow rate. The outflow is typically mainly H$^+$ with significant fractions of O$^+$ and He$^+$. Overall, the average dayside outflow rate is of the same order of magnitude as the high latitude upflow rate, $10^{26}$ ions/s. 

\begin{table}[b]
   \caption{Examples of statistical studies of ion upflow at high latitudes and ion outflow through the magnetopause at low latitudes. On Cluster the spacecraft wake technique was used to cover low energies at high latitudes and a combination of techniques was used at low latitudes; on DE, Akebono and Polar (TIDE and TIMAS instruments) onboard particle detectors were used. The Magnetospheric Plasma Analyzers (MPA) are located on multiple geostationary satellites, while IMAGE instruments used a remote sensing technique. Spacecraft charging is a smaller problem for heavier ions since for a given velocity they have higher energy, so O$^+$ ions can be observed by onboard detectors on Cluster. Typical ion upflow/outflow rates are given. Rates can vary at least over an order of magnitude, in particular for heavier ions such as O$^+$ at increased solar EUV and geophysical activity. Conditions for the individual studies are given in the references. At the magnetopause ions are typically H$^+$ with significant fractions of O$^+$ and He$^+$. Overall a typical outflow rate is 10$^{26}$ ions/s.
   \\}\label{t1}
   \centering
   \begin{tabular}{lrrr}
	\hline
 \multicolumn{4}{l}{\textbf{High latitude: Cusp/cleft, Polar Cap, Auroral Region}}\\
  \underline{Spacecraft}	&	\underline{Nominal energy range (eV)}	&	\underline{Altitude (R$_E$)}	&	\underline{Upflow rate (10$^{26}$ ions/s); ion species}	\\
  Cluster$^{1}$			&	0 - 60						&	5 - 20				&	0.6 - 2.4 (mainly H$^+$)				\\
  Polar/TIDE$^{2}$			&	<1 - 100						&	8					&	1.3 (mainly H$^+$)						\\
  Polar/TIDE$^{3}$			&	<1 - 450					&	0.8					&	1.7 (mainly H$^+$)							\\
  Akebono$^{4}$			&	<1 - 70						&	1 - 1.5				&	 0.2 - 2(H$^+$ and O$^+$)				\\
  DE$^{5}$				&	10 - 17,000					&	2.5-3.7					&	0.2 - 2 (H$^+$ and O$^+$)				\\
  Polar/TIMAS$^{6}$			&	15 - 33,000					&	0.8				&	0.08 (H$^+$ and O$^+$)				\\
     Cluster/CODIF$^{7}$			&	25 - 38,000						&	10 - 15				&	0.1 - 2.2 (O$^+$)				\\
	\hline
     \multicolumn{4}{l}{\textbf{Low latitude: Magnetopause (plumes)}}\\
  \underline{Spacecraft}	&	\underline{Nominal energy range (eV)}	&		&	\underline{Outflow rate (10$^{26}$ ions/s)	} \\
  Cluster$^{8}$			&	0 - 1000						&						&	1 - 10				\\
  MPA$^{9}$	&	1 - 40,000					&						&	2					\\
  IMAGE$^{10}$			&	wide range							&						&	3.8 - 21						\\
	\hline
      \multicolumn{4}{l}{\textbf{Low latitude: Magnetopause (wind)}}\\
  \underline{Spacecraft}	&	\underline{Nominal energy range (eV)}	&		&	\underline{Outflow rate (10$^{26}$ ions/s)	 } \\
  Cluster$^{8}$		&	0 - 1000						&						&	0.1 - 1				\\
	\hline
    \multicolumn{4}{l}{
    $^{1}$\citeA{Andre2015a}, 
    $^{2}$\citeA{Su1998},
    $^{3}$\citeA{Huddleston2005}, 
    $^{4}$\citeA{Cully2003},} \\
    \multicolumn{4}{l}{
    $^{5}$\citeA{YauAndre1997}, 
    $^{6}$\citeA{Peterson2006, Peterson2008}, 
    $^{7}$\citeA{Slapak2017}, 
    $^{8}$\citeA{AndreCully2012}, } \\    
    \multicolumn{4}{l}{$^{9}$Magnetospheric Plasma Analyzers \citeA{BorovskyDenton2008},
  $^{10}$ \citeA{SpasojevicSandel2010}, 
	} \\
   \end{tabular}
\end{table}

All ions are cold in the collisional ionosphere. In the high latitude auroral region, some ions are energized to several hundred eV or a few keV energies within a few thousand km altitude, by waves (perpendicular to \textbf{B}) and quasi-static auroral electric fields (parallel to \textbf{B}), forming ion conics and beams, respectively, e.g. \citeA{AndreYau1997,Moore2007}. In the polar cap, magnetotail lobes, ions may stay at low energies to much higher altitudes but can gradually be energized by waves and centrifugal acceleration \cite{Nilsson2013}. A mixture of cold and hot ions of ionospheric origin can occur together, since the ions along different paths may experience different heating mechanisms and respond in different ways due to their different masses, e.g. \citeA{Nilsson2012, Liao2012}.

Most of the upflowing ions at high latitudes reaching altitudes of a few thousand kilometers do not return to the collisional ionosphere. Comparing the outflow rate data in Table 1 with precipitating ions at tens of eV and higher (5-8 $10^{24}$ ions/s) observed on the DMSP satellites below a thousand kilometers \cite{Newell2009} implies that only a small fraction of these upflowing ions return. This is not surprising since their velocity vector would have to be within the atmospheric loss-cone in order for them not to mirror above the collisional region.   

Upflowing cold (or gradually heated) ions in the magnetotail lobes may be an important source for the plasma sheet, at least sometimes. Rather than escaping down the geomagnetic tail these ions for are convected to the plasma sheet \cite{Huddleston2005, Haaland2012, LiK2013} under many conditions. Some of the ions are then heated before they leave the magnetosphere, some may leave the magnetosphere as rather cold ions. Ions of ionospheric origin transported toward the nightside and then brought back to the dayside magnetosphere by {\textbf{E} $\times$ \textbf{B}} convection can form the so-called warm plasma cloak. These ions are energized to an intermediate energy (a few eV to hundreds of eV), which is greater than the energy of directly upflowing ions and less than that of the plasma sheet or ring current ions \cite{Chappell2008, Lee2016}. When reaching a reconnection region at the magnetopause these ions may be heated to higher energies, or stay at rather low thermal velocities when leaving the magnetosphere. 

The outflow of originally cold ionospheric ions is a significant part of the atmospheric outflow leaving planet Earth (also including neutrals) \cite{Engwall2009a, Andre2015b}. 

\section{Effects at Small Scales}

Cold ions are present in large regions and can affect kinetic plasma processes in small regions which then act back on large regions. Increasing the (mass) density, for example, by increasing the density of cold ionospheric ions, will decrease the Alfv\'en velocity and hence the reconnection rate. At the magnetopause, effects of cold ions are more likely with a compressed magnetosphere. This can be significant for high-density plasmaspheric plumes, but is usually not a major effect \cite{Lee2016, Fuselier2017}. 

\begin{figure*}
	\includegraphics[width=70 mm]{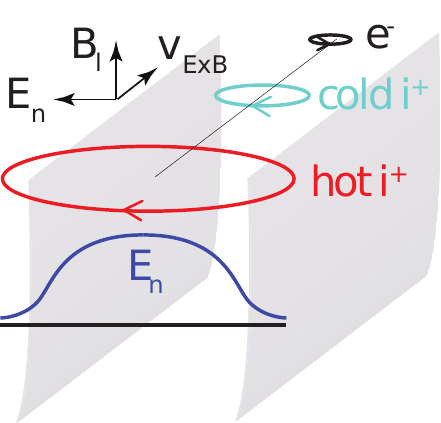}
    \caption{Sketch of a magnetic reconnection separatrix with an electric field in the normal direction, and the different length-scales associated with hot (keV) and cold (eV) ions, and electrons, respectively. The hot ions are not magnetized, while the cold (eV) ions and the electrons are magnetized (see \citeA{Toledo-Redondo2015}).
    }
    \label{fig:figure-Sketch_Andre_GRL_2016}
\end{figure*}

An interesting effect in small regions (order of gyroradius) is that cold ions introduce a new length-scale. The cold ions have a gyroradius between the gyroradii of hot (keV) ions and electrons, which have much lower mass. This is somewhat similar to introducing a new length-scale by introducing heavy ions with a gyroradius larger than other species. An important example is the small scales associated with magnetic reconnection, including the thin separatrix regions extending out from the diffusion region. Figure~\ref{fig:figure-Sketch_Andre_GRL_2016} shows a sketch of a magnetic separatrix region, and the multiple length scales involved. Here the hot ions are not magnetized, while both cold ions and electrons are magnetized. In this thin boundary cold ions {\textbf{E} $\times$ \textbf{B}} drift together with electrons and reduce the Hall current. 

\begin{figure*}
	\includegraphics[width= 80 mm]{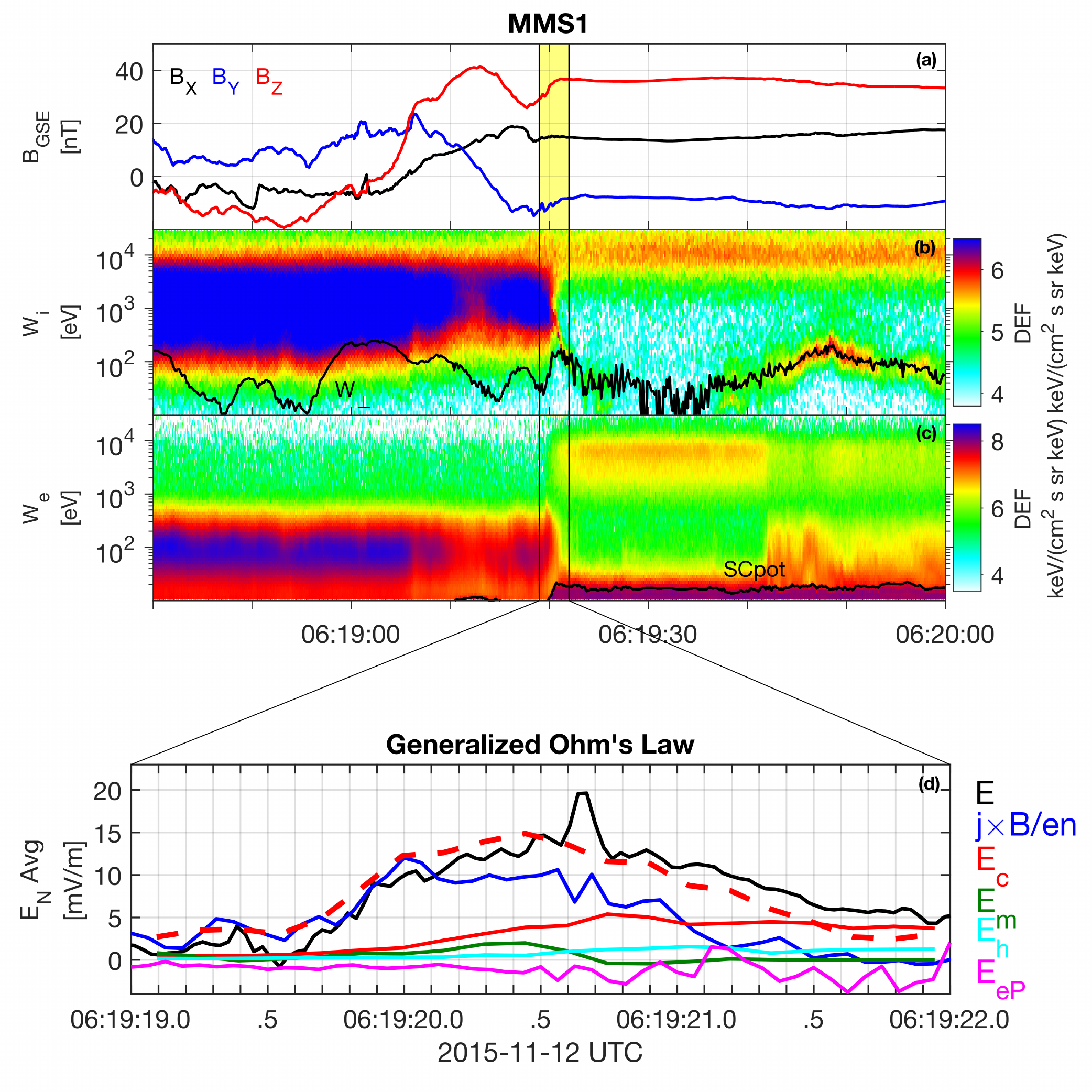}
    \caption{Crossing from the magnetosheath to the magnetosphere of 
    a magnetic reconnection separatrix observed by the MMS spacecraft. Panel a) MMS 1 magnetic field in GSE coordinates,  b) ion differential energy flux and average energy obtained in the perpendicular direction c) electron differential energy flux and spacecraft potential d) components of the generalized Ohms law averaged over all MMS spacecraft, in the direction normal to the separtarix current sheet: the electric field (black) is balanced by a combination of the Hall term {\textbf{j} $\times$ \textbf{B}}/en (blue), {\textbf{E} $\times$ \textbf{B}} drifting cold ions E\textsubscript{c} (red) with some contribution from the gradient of the electron pressure E\textsubscript{eP} (purple), the sum is shown as a dashed line (see \citeA{Andre2016}).
    }
    \label{fig:figure-OhmsLaw_Andre_GRL_2016}
\end{figure*}

A crossing of a magnetopause reconnection separatrix by the four MMS spacecraft is shown in Figure~\ref{fig:figure-OhmsLaw_Andre_GRL_2016}. Here cold ions with an {\textbf{E} $\times$ \textbf{B}} drift energy of 20-200 eV are found on the magnetospheric (right) side, panel b). The electric field normal to the thin separatrix current sheet (black line, panel d) is in the generalized Ohms law balanced by the reduced {\textbf{v} $\times$ \textbf{B}} Hall current term (blue), the term representing drifting cold ions (red) with some contribution from the gradient of the electron pressure (purple) \cite{Andre2016}. Without the cold ions the normal electric field is often balanced by the Hall term, e.g. \citeA{Khotyaintsev2006}. The situation with cold ions changing the Hall physics of reconnection is common at the magnetopause. This is important for the kinetic physics, and when using observations of currents to investigate reconnection and other thin boundaries. The effects of cold ionospheric ions on kinetic plasma physics in thin layers have been investigated using multi-spacecraft observations by Cluster and MMS at the magnetopause \cite{Toledo-Redondo2015, Andre2016, Toledo-Redondo2016a, Toledo-Redondo2017} by MMS in the magnetotail \cite{Alm2018}, and using simulations \cite{Dargent2017,Toledo-Redondo2018}.

\begin{figure*}
	\includegraphics[width=120 mm]{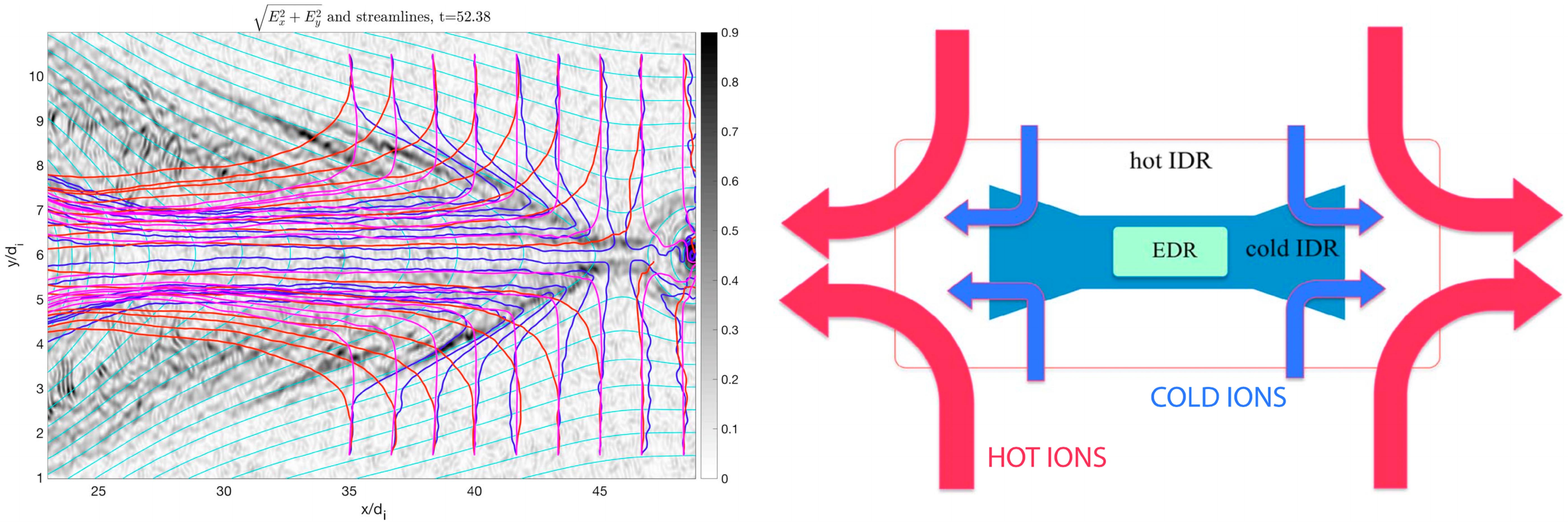}
    \caption{(Left) Simulation of magnetic reconnection, with separatrices and multiple diffusion regions, including hot and cold ions, and electrons. Streamlines of hot and cold ions and electrons (red, magenta and blue), and magnitude of the electric field (grey scale). (Right) Sketch illustrating the diffusion region layout when cold ions are present in symmetric magnetic reconnection, from \citeA{Divin2016}
    }
    \label{fig:figure-Simulation_Divin_JGR_2016_low}
\end{figure*}

Close to the reconnection X-line an additional diffusion region is created (hot and cold ions, electrons, total of three). This is similar to the effect of heavier ions with larger gyro radius \cite{Liu2015}. This is indicated in MMS observations, \cite{Toledo-Redondo2016b}, and is found in simulations on kinetic scales, \cite{Divin2016}, Fig.~\ref{fig:figure-Simulation_Divin_JGR_2016_low}.

Cold ions can be heated close to the reconnecting magnetopause boundary up to several hundred eV, as shown in Fig. 5. Strong heating occurs inside the magnetospheric sepatratrix region, marked in yellow, where waves close to the ion gyrofrequency and sharp spatial gradients in the electric field are observed. A significant fraction of the energy going into ion heating can go into heating of initially cold ions \cite{Toledo-Redondo2016b, Toledo-Redondo2017}. Cold ions may be heated when they hit the separatrix region, but sometimes cold ions can also be found deep into the outflow jet \cite{LiW2017}.

\begin{figure*}
\includegraphics[width=120 mm]{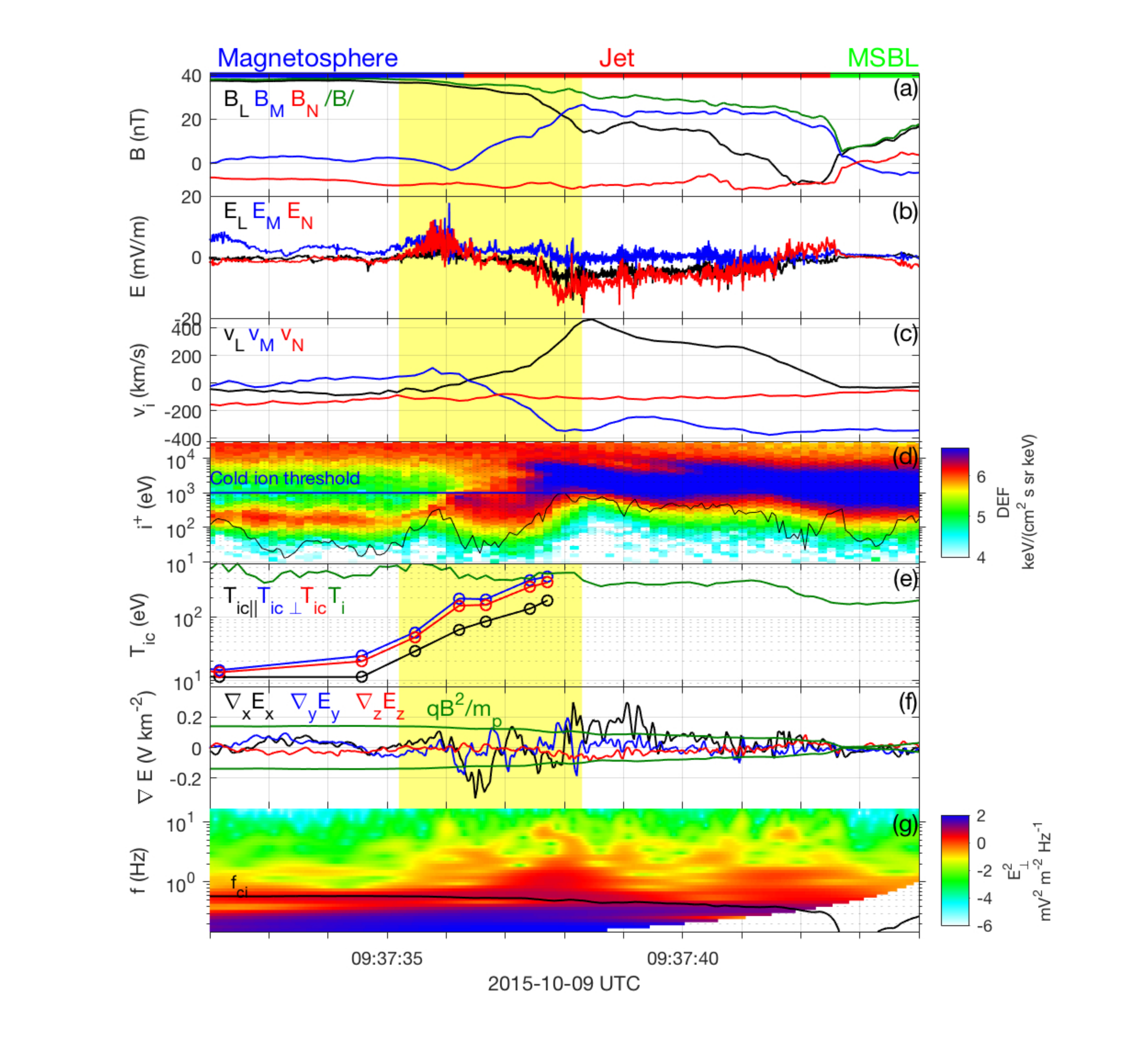}
    \caption{MMS measurement of cold ion heating in the dayside magnetospheric separatrix region of magnetic reconnection. Panel a) Magnetic field; b) electric field; c) ion velocity; d) Omnidirectional ion differential energy flux; e) cold ion temperatures; f) electric field gradient (used to estimate possible ion heating); g) electric field wave spectrogram (used to estimate possible ion heating). The cold ion temperature is below a few tens of eV in the inflow region (before 09:37:35), and increases to a few hundred eV inside the magnetospheric separatrix region (marked in yellow). The ions are heated by electric field gradients and waves (panels b, f and g). 
    (From \citeA{Toledo-Redondo2017})}
    \label{fig:figure-Event_Sergio_JGR_2017}
\end{figure*}

Cold ions affect the kinetic plasma instabilities that can occur. For example, not only density gradients but also the relative drift between (magnetized) cold ions and (unmagnetized) hot ions can generate lower-hybrid waves \cite{Graham2017}. These waves act back on the particle populations. For example, the waves can heat the initially cold ions, and can smooth density gradients.




\section{Space Weather Effects}

Effects of ionospheric ions include changes in the Alfv\'en velocity (affecting energy transport and the reconnection rate). Effects of cold ionospheric ions also include changes of the kinetic plasma physics (including reconnection, at the magnetopause and the magnetotail). Thus, ionospheric ions affect energy transport from the solar wind to the magnetosphere, and energy storage and release in the magnetospheric system. Strong space weather events are often associated with increased solar wind pressure leading to a compressed dayside magnetosphere and more cold ions from the plasmasphere at the magnetopause, potentially affecting the reconnection rate. The effects of ionospheric ions should be included in future space weather studies.     

\section{Discussion and Open Questions}

In the case of cold low-energy ions statistical studies can be very dependent on the measurement techniques and results may at first seem contradictory. Some studies may emphasize ions at rather high energies that can be detected also onboard a charged spacecraft, while another method indicates that the plasma is dominated by low-energy ions at the same time. Both studies can be correct, since a low-energy ion population can be detected with one method, while the ions with higher energies can be detected with another method. As an example, the spacecraft wake technique can estimate the flux of low-energy ions also in the presence of a population of ions at higher energies. As long as the low-energy population dominates the density a wake will form, indicating the presence of cold ions. The ions at higher energy can still reach the spacecraft and be detected, without destroying the wake. Hence, two ion populations at different energies may exist at the same time, and may individually be detected with different methods.


Concerning open questions, it is not clear how much low-energy ionospheric ions contribute to some large-scale structures in the magnetosphere by directly supplying material. For example, are upflowing ionospheric ions, most of the time, essential for forming the plasma sheet? In addition, it is not clear how important cold ions of ionospheric origin are for small-scale kinetic plasma physics, which then will influence large-scale structures. Cold ions in thin current sheets, such as reconnection separatrices, can reduce the Hall current, modify the terms in Ohms law, and introduce new wave instabilities. This will act back on the mechanisms of magnetic reconnection. However, it is not clear how much modifications at the kinetic scale will influence the reconnection rate. 
%

\section{Summary}

Cold (eV) ions dominate the number density of most of the volume of the magnetosphere during most of the time. This affects large scales, including the Alfv\'en velocity and thus energy transport with waves and the magnetic reconnection rate. This also affects small scale kinetic plasma physics, including the Hall physics and wave instabilities associated with magnetic reconnection. Open questions include how ionospheric ions contribute to large-scale structures such as the plasma sheet, and how effects of low-energy ions on kinetic scales affect the large-scale reconnection rate. Effects of ionospheric plasma should be included in investigations at all scales in the magnetosphere and can be essential for studies of space weather.

\acknowledgments
We acknowledge support to MA from the Swedish National Space Agency under contracts SNSB 176/15 and 287/16, and from the Natural Science and Engineering Research Council to AWY under Discovery Grant RGPIN/06069-2014.
We also acknowledge support from the ISSI international team \textit{Cold plasma of ionospheric origin at the Earths magnetosphere}.






%
%
%
%
%
%

\bibliography{Cold_ions.bib}

%
%
%
%






\end{document}